%
%

\input harvmac

\noblackbox

\def\cO{{\cal O}}

\let\includefigures=\iftrue
\let\useblackboard=\iftrue
\newfam\black

\includefigures
\message{If you do not have epsf.tex (to include figures),}
\message{change the option at the top of the tex file.}
\def\figin{\epsfcheck\figin}\def\figins{\epsfcheck\figins}
\def\epsfcheck{\ifx\epsfbox\UnDeFiNeD
\message{(NO epsf.tex, FIGURES WILL BE IGNORED)}
\gdef\figin##1{\vskip2in}\gdef\figins##1{\hskip.5in}
\else\message{(FIGURES WILL BE INCLUDED)}%
\gdef\figin##1{##1}\gdef\figins##1{##1}\fi}
\def\DefWarn#1{}
\def\figinsert{\goodbreak\midinsert}
\def\ifig#1#2#3{\DefWarn#1\xdef#1{Fig.~\the\figno}
\writedef{#1\leftbracket Fig.\noexpand~\the\figno}%
\figinsert\figin{\centerline{#3}}\medskip\centerline{\vbox{
\baselineskip12pt\advance\hsize by -1truein
\noindent\footnotefont{\bf Fig.~\the\figno:} #2}}
\bigskip\endinsert\global\advance\figno by1}
\else
\def\ifig#1#2#3{\xdef#1{Fig.~\the\figno}
\writedef{#1\leftbracket Fig.\noexpand~\the\figno}%
\global\advance\figno by1} \fi

\def\doublefig#1#2#3#4{\DefWarn#1\xdef#1{Fig.~\the\figno}
\writedef{#1\leftbracket Fig.\noexpand~\the\figno}%
\figinsert\figin{\centerline{#3\hskip1.0cm#4}}\medskip\centerline{\vbox{
\baselineskip12pt\advance\hsize by -1truein
\noindent\footnotefont{\bf Fig.~\the\figno:} #2}}
\bigskip\endinsert\global\advance\figno by1}

\useblackboard
\message{If you do not have msbm (blackboard bold) fonts,}
\message{change the option at the top of the tex file.}
\font\blackboard=msbm10 scaled \magstep1 \font\blackboards=msbm7
\font\blackboardss=msbm5 \textfont\black=\blackboard
\scriptfont\black=\blackboards \scriptscriptfont\black=\blackboardss

\else

\fi
%

\def\yboxit#1#2{\vbox{\hrule height #1 \hbox{\vrule width #1
\vbox{#2}\vrule width #1 }\hrule height #1 }}
\def\fillbox#1{\hbox to #1{\vbox to #1{\vfil}\hfil}}
\def\ybox{{\lower 1.3pt \yboxit{0.4pt}{\fillbox{8pt}}\hskip-0.2pt}}
%
%



\def\comments#1{}

\def\Tr{{{\rm Tr~ }}}

\def\vev#1{\langle{#1}\rangle}



\def\II{\relax{I\kern-.10em I}}

\def\IZ{\relax{\rm Z\kern-.34em Z}}
\def\IB{\relax{\rm I\kern-.18em B}}
\def\IC{{\relax\hbox{$\inbar\kern-.3em{\rm C}$}}}
\def\ID{\relax{\rm I\kern-.18em D}}
\def\IE{\relax{\rm I\kern-.18em E}}
\def\IF{\relax{\rm I\kern-.18em F}}
\def\IG{\relax\hbox{$\inbar\kern-.3em{\rm G}$}}
\def\IGa{\relax\hbox{${\rm I}\kern-.18em\Gamma$}}
\def\IH{\relax{\rm I\kern-.18em H}}
\def\II{\relax{\rm I\kern-.18em I}}
\def\IK{\relax{\rm I\kern-.18em K}}
\def\IP{\relax{\rm I\kern-.18em P}}

%

\def\inbar{\,\vrule height1.5ex width.4pt depth0pt}

\def\IR{\relax{\rm I\kern-.18em R}}

\def\simgt{\hskip0.05in\relax{
\raise3.0pt\hbox{ $>$ {\lower5.0pt\hbox{\kern-1.05em $\sim$}} }}
\hskip0.05in}

%


%

\def\lp10{\ell_p^{10}}
\def\lp11{\ell_p^{11}}
\def\R11{R_{11}}

\def\frac#1#2{{#1 \over #2}}



\newdimen\tableauside\tableauside=1.0ex
\newdimen\tableaurule\tableaurule=0.4pt
\newdimen\tableaustep
\def\phantomhrule#1{\hbox{\vbox to0pt{\hrule height\tableaurule width#1\vss}}}
\def\phantomvrule#1{\vbox{\hbox to0pt{\vrule width\tableaurule height#1\hss}}}
\def\sqr{\vbox{%
  \phantomhrule\tableaustep
  \hbox{\phantomvrule\tableaustep\kern\tableaustep\phantomvrule\tableaustep}%
  \hbox{\vbox{\phantomhrule\tableauside}\kern-\tableaurule}}}
\def\squares#1{\hbox{\count0=#1\noindent\loop\sqr
  \advance\count0 by-1 \ifnum\count0>0\repeat}}
\def\tableau#1{\vcenter{\offinterlineskip
  \tableaustep=\tableauside\advance\tableaustep by-\tableaurule
  \kern\normallineskip\hbox
    {\kern\normallineskip\vbox
      {\gettableau#1 0 }%
     \kern\normallineskip\kern\tableaurule}%
  \kern\normallineskip\kern\tableaurule}}
\def\gettableau#1 {\ifnum#1=0\let\next=\null\else
  \squares{#1}\let\next=\gettableau\fi\next}

\tableauside=1.0ex \tableaurule=0.4pt


 %
 %
 \def\eqnn#1{\xdef #1{(\secsym\the\meqno)}\writedef{#1\leftbracket#1}%
 \global\advance\meqno by1\wrlabeL#1}
 \def\eqna#1{\xdef #1##1{\hbox{$(\secsym\the\meqno##1)$}}
 \writedef{#1\numbersign1\leftbracket#1{\numbersign1}}%
 \global\advance\meqno by1\wrlabeL{#1$\{\}$}}
 \def\eqn#1#2{\xdef #1{(\secsym\the\meqno)}\writedef{#1\leftbracket#1}%
 \global\advance\meqno by1$$#2\eqno#1\eqlabeL#1$$}

\global\newcount\itemno \global\itemno=0

\def\itemaut#1{\global\advance\itemno by1\noindent\item{\the\itemno.}#1}



\hyphenation{Di-men-sion-al}




\lref\HellermanNX{
  S.~Hellerman and I.~Swanson,
  ``Cosmological solutions of supercritical string theory,''
  arXiv:hep-th/0611317.
}

\lref\hellliu{
  S.~Hellerman and X.~Liu,
  ``Dynamical dimension change in supercritical string theory,''
  arXiv:hep-th/0409071.
}

\lref\mss{
  E.~Silverstein,
  ``(A)dS backgrounds from asymmetric orientifolds,''
  contribution to Strings 2001 [hep-th/0106209];
  A.~Maloney, E.~Silverstein and A.~Strominger,
  ``De Sitter space in noncritical string theory,''
  in {\it Cambridge 2002: The future of theoretical physics and
  cosmology}, 570-591
 [hep-th/0205316].}

\lref\HellermanZM{
  S.~Hellerman,
  ``On the landscape of superstring theory in $D > 10$,''
  arXiv:hep-th/0405041.
}

\lref\McGreevyCI{
  J.~McGreevy and E.~Silverstein,
  ``The tachyon at the end of the universe,''
  JHEP {\bf 0508}, 090 (2005)
  [arXiv:hep-th/0506130].
}

\lref\SuyamaWD{
  T.~Suyama,
  ``Closed string tachyon condensation in supercritical strings and RG flows,''
  JHEP {\bf 0603}, 095 (2006)
  [arXiv:hep-th/0510174].
}

\lref\FreedmanWX{
  D.~Z.~Freedman, M.~Headrick and A.~Lawrence,
  ``On closed string tachyon dynamics,''
  Phys.\ Rev.\ D {\bf 73}, 066015 (2006)
  [arXiv:hep-th/0510126].
}

\lref\TseytlinYE{
  A.~A.~Tseytlin,
   ``Cosmological solutions with dilaton and maximally symmetric space in string
  theory,''
  Int.\ J.\ Mod.\ Phys.\ D {\bf 1}, 223 (1992)
  [arXiv:hep-th/9203033].
}

\lref\SchmidhuberBV{
  C.~Schmidhuber and A.~A.~Tseytlin,
  ``On string cosmology and the RG flow in 2-d field theory,''
  Nucl.\ Phys.\ B {\bf 426}, 187 (1994)
  [arXiv:hep-th/9404180].
}

\lref\mutation{
  E.~Silverstein,
  ``Dimensional mutation and spacelike singularities,''
  Phys.\ Rev.\ D {\bf 73}, 086004 (2006)
  [arXiv:hep-th/0510044].
}

\lref\growthnotes{J. McGreevy, E. Silverstein, D. Starr, ``New
Dimensions for Wound Strings", to appear.}

\lref\Dduality{D. Green, A. Lawrence, J. McGreevy, D. Morrison, E.
Silverstein, ``Dimensional Duality", in progress.}

\lref\PolyakovRD{
  A.~M.~Polyakov,
  ``Quantum Geometry Of Bosonic Strings,''
  Phys.\ Lett.\ B {\bf 103}, 207 (1981).
}

\lref\CooperVG{
  A.~R.~Cooper, L.~Susskind and L.~Thorlacius,
  ``Two-dimensional quantum cosmology,''
  Nucl.\ Phys.\ B {\bf 363}, 132 (1991).
}

\lref\KutasovUA{
  D.~Kutasov and N.~Seiberg,
  ``Noncritical Superstrings,''
  Phys.\ Lett.\ B {\bf 251}, 67 (1990).
}

\lref\WittenYU{
  E.~Witten,
  ``On the conformal field theory of the Higgs branch,''
  JHEP {\bf 9707}, 003 (1997)
  [arXiv:hep-th/9707093].
}

\lref\AharonyTH{
  O.~Aharony, M.~Berkooz, S.~Kachru, N.~Seiberg and E.~Silverstein,
  ``Matrix description of interacting theories in six dimensions,''
  Adv.\ Theor.\ Math.\ Phys.\  {\bf 1}, 148 (1998)
  [arXiv:hep-th/9707079].
}

\lref\DiaconescuGU{
  D.~E.~Diaconescu and N.~Seiberg,
  ``The Coulomb branch of (4,4) supersymmetric field theories in two
  dimensions,''
  JHEP {\bf 9707}, 001 (1997)
  [arXiv:hep-th/9707158].
}

\lref\StromTak{
  A.~Strominger and T.~Takayanagi,
``Correlators in timelike bulk Liouville theory,''
  Adv.\ Theor.\ Math.\ Phys.\  {\bf 7}, 369 (2003)
  [arXiv:hep-th/0303221].
}

\lref\MukherjiJI{
  S.~Mukherji,
  ``On The Interpolating Solutions Of String In Different Backgrounds,''
  Mod.\ Phys.\ Lett.\ A {\bf 7}, 1361 (1992)
  [arXiv:hep-th/9203010].
}

\lref\EllisEH{
  J.~R.~Ellis, N.~E.~Mavromatos and D.~V.~Nanopoulos,
  ``String Theory Modifies Quantum Mechanics,''
  Phys.\ Lett.\ B {\bf 293}, 37 (1992)
  [arXiv:hep-th/9207103].
}

\lref\EllisBA{
  J.~R.~Ellis, N.~E.~Mavromatos and D.~V.~Nanopoulos,
  ``A Liouville String Approach To Microscopic Time And Cosmology,''
  arXiv:hep-th/9311148, Lectures given at International Workshop on
Recent Advances in the Superworld, Woodlands, TX, 13-16 Apr 1993,
published in Woodlands Superworld 1993:3-26.
}

\lref\PolchinskiFN{
  J.~Polchinski,
  ``A two-dimensional model for quantum gravity,''
  Nucl.\ Phys.\ B {\bf 324}, 123 (1989).
}

\lref\kutsei{
  D.~Kutasov and N.~Seiberg,
   ``Number Of Degrees Of Freedom, Density Of States And Tachyons In String
  Theory And Cft,''
  Nucl.\ Phys.\ B {\bf 358}, 600 (1991).
}

\lref\AharonyDW{
  O.~Aharony and M.~Berkooz,
  ``IR dynamics of d = 2, N = (4,4) gauge theories and DLCQ of 'little  string
  theories',''
  JHEP {\bf 9910}, 030 (1999)
  [arXiv:hep-th/9909101].
}

\lref\BirrellIX{
  N.~D.~Birrell and P.~C.~W.~Davies,
  ``Quantum Fields In Curved Space,''
  Cambridge University Press.
}

\lref\StromingerPC{
  A.~Strominger,
  ``Open string creation by S-branes,''
  arXiv:hep-th/0209090.
}

\lref\MyersFV{
  R.~C.~Myers,
  ``New Dimensions For Old Strings,''
  Phys.\ Lett.\ B {\bf 199}, 371 (1987).
}

\lref\deAlwisPR{
  S.~P.~de Alwis, J.~Polchinski and R.~Schimmrigk,
  ``Heterotic strings with tree level cosmological constant,''
  Phys.\ Lett.\ B {\bf 218}, 449 (1989).
}

\lref\AntoniadisAA{
  I.~Antoniadis, C.~Bachas, J.~R.~Ellis and D.~V.~Nanopoulos,
  ``Cosmological string theories and discrete inflation,''
  Phys.\ Lett.\  B {\bf 211}, 393 (1988).
}

\lref\AntoniadisVI{
  I.~Antoniadis, C.~Bachas, J.~R.~Ellis and D.~V.~Nanopoulos,
  ``An expanding universe in string theory,''
  Nucl.\ Phys.\ B {\bf 328}, 117 (1989).
}

\lref\AntoniadisUU{
  I.~Antoniadis, C.~Bachas, J.~R.~Ellis and D.~V.~Nanopoulos,
  ``Comments on cosmological string solutions,''
  Phys.\ Lett.\ B {\bf 257}, 278 (1991).
}

\lref\AharonyCX{
  O.~Aharony, M.~Fabinger, G.~T.~Horowitz and E.~Silverstein,
  ``Clean time-dependent string backgrounds from bubble baths,''
  JHEP {\bf 0207}, 007 (2002)
  [arXiv:hep-th/0204158].
}

\lref\PolyakovRD{
  A.~M.~Polyakov,
  ``Quantum geometry of bosonic strings,''
  Phys.\ Lett.\ B {\bf 103}, 207 (1981).
}

\lref\PolyakovRE{
  A.~M.~Polyakov,
  ``Quantum geometry of fermionic strings,''
  Phys.\ Lett.\ B {\bf 103}, 211 (1981).
}

\lref\AdamsRB{
  A.~Adams, X.~Liu, J.~McGreevy, A.~Saltman and E.~Silverstein,
  ``Things fall apart: Topology change from winding tachyons,''
  JHEP {\bf 0510}, 033 (2005)
  [arXiv:hep-th/0502021].
}

\lref\EllisQA{
  J.~R.~Ellis, N.~E.~Mavromatos, D.~V.~Nanopoulos and M.~Westmuckett,
  ``Liouville cosmology at zero and finite temperatures,''
  Int.\ J.\ Mod.\ Phys.\ A {\bf 21}, 1379 (2006)
  [arXiv:gr-qc/0508105].
}

\lref\AlexandreAT{
  J.~Alexandre, J.~Ellis and N.~E.~Mavromatos,
  ``Non-perturbative formulation of non-critical string models,''
  arXiv:hep-th/0611228.
}


\Title{\vbox{\baselineskip12pt\hbox{hep-th/0612031}
\hbox{SU-ITP-06/32}\hbox{SLAC-PUB-12243}\hbox{WIS/18/06-NOV-DPP}}}
{\vbox{ \centerline{} \centerline{Supercritical Stability,
Transitions and (Pseudo)tachyons}\centerline{}
%
%
%
%
%
%
%
%
%
}}
\bigskip
\centerline{Ofer Aharony$^{a,b}$, and Eva Silverstein$^{c,d}$}
\bigskip
\centerline{$^a${\it Department of Particle Physics,
Weizmann Institute of Science, Rehovot 76100, Israel.}}
\bigskip
\centerline{$^b${\it School of Natural Sciences, Institute for
Advanced Study, Princeton, NJ 08540, USA.}}
\bigskip
\centerline{$^c${\it SLAC and Department of Physics, Stanford
University, Stanford, CA 94305-4060, USA.}}
\bigskip
\centerline{$^d${\it Kavli Institute for Theoretical Physics,
University of California, Santa Barbara, CA 93106-4030, USA.}}
\bigskip
\bigskip
\noindent
Highly supercritical strings ($c\gg 15$) with a time-like linear
dilaton provide a large class of solutions to string theory, in
which closed string tachyon condensation is under control (and
follows the worldsheet renormalization group flow). In this note
we analyze the late-time stability of
such backgrounds, including transitions between them. The large
friction introduced by the rolling dilaton and the rapid decrease of
the string coupling suppress the back-reaction of naive
instabilities. In particular, although the graviton, dilaton, and
other light fields have negative effective mass squared in the
linear dilaton background, the decaying string coupling ensures that
their condensation does not cause large back-reaction. Similarly,
the copious particles produced in transitions between highly
supercritical theories do not back-react significantly on the
solution.  We discuss these features also in a somewhat more general
class of time-dependent backgrounds with stable late-time
asymptotics.

\bigskip
\Date{December 2006}



\newsec{Introduction}

It is of interest to understand cosmological solutions of string
theory. Most solutions of general relativity coupled to quantum
field theory evolve non-trivially with time, leading to a weakly
coupled description only (at best) at asymptotically late times (or
only at early times). This general class of backgrounds includes
Calabi-Yau manifolds with flux\foot{Except when these lead to
an anti-de Sitter solution.}, curved target spaces, and many
others.

For a variety of applications, one might like to understand the
physics of the perturbation spectrum about such backgrounds in the
easily controlled weakly coupled regime. This is important for a
stability analysis, as well as for calculating the particle content
and density perturbations in cosmological solutions.  Moreover, in
perturbative string theory, the infrared perturbation spectrum is
related by modular invariance to microphysical information, i.e. to
the high-energy density of single-string states. More generally, one
would like to understand the transitions connecting different
tractable limits, which in some cases may arise as closed string
tachyon condensation processes.

Perhaps the simplest background with a weakly coupled future
asymptotic region is the time-like linear dilaton solution of
supercritical superstring theory
\refs{\AntoniadisAA
\PolchinskiFN\AntoniadisVI\AntoniadisUU\CooperVG
\MukherjiJI\TseytlinYE
\SchmidhuberBV\EllisEH\EllisBA\MyersFV-\deAlwisPR}\foot{For
more recent discussions, see \refs{\EllisQA,\AlexandreAT} and
references therein.}, formulated
in $d=10 + 2Q^2$ dimensions ($Q>0$). In the string frame this
background is flat. In the Einstein frame its metric is
\eqn\SCLDE{ds^2_E=-dt^2+{4Q^2\over{(d-2)^2}}t^2d\vec x^2,}
with the string coupling
\eqn\gE{g_s(t)=\left({d-2\over 2Qt}\right)^{(d-2)/2}.}
We will refer to this as the supercritical linear dilaton (SCLD)
phase.  As shown in early works, it constitutes an exact classical
solution to string theory, including all $\alpha'$ corrections.
These backgrounds are strongly coupled at early times, so in order
to completely define them we need to ``cap off'' the strong coupling
region. One approach to this is to imagine tunneling into the SCLD
phase from some meta-stable background (perhaps during eternal
inflation) such as the background considered in \mss\ (and
generalizations), though this does not address the physics
arbitrarily far back in time. The details of this capping will not
be important for most of our considerations.

In this note, we analyze the stability and late-time perturbation
spectrum of these theories and of closely related solutions
describing transitions between them.  Some of our considerations
apply more generally as we will indicate as we go along.
We begin in \S2 by reviewing the properties of SCLD backgrounds and
the fluctuations around them, and the fact that in the large $Q$ limit,
the tachyon condensation process in SCLD backgrounds follows the
renormalization group flow on the worldsheet.

The analysis of the SCLD spectrum leads to what we call
{\it pseudotachyons}.  These are mode
solutions which do not oscillate in time, and which naively cause an IR
divergence in the perturbative string partition function at
one-loop. In \S3 we discuss these modes, and we argue that they do
not actually cause an instability of the SCLD backgrounds, since their
condensation does not cause a large back-reaction (a condition we
will quantify).

This phenomenon appears much more generally in weakly coupled
asymptotic regimes of general relativity and string theory; a
prototypical example is the perturbations produced during
inflationary expansion periods. Modular invariance relates this IR
divergence to the supercritical effective central charge arising for
$d>10$. Since there are many similar backgrounds with
pseudotachyonic perturbations, an interesting corollary is that any
consistent perturbative string background with pseudotachyons is
effectively supercritical, even if extra dimensions are not
specified explicitly.  An interesting class of examples of this is
compact negatively curved target spaces, where the requisite
effective central charge arises from winding modes
\refs{\mutation\growthnotes-\Dduality}.

In \S4 we discuss the particle production during tachyon
condensation processes in SCLD backgrounds. We argue that even
though there is large particle production (whose details depend on
the initial conditions for the SCLD phase), the produced particles
do not back-react strongly on the SCLD background, so they do not
change the classical fact that the tachyon condensation process
follows the worldsheet RG flow.  We end in \S5 with a summary of our
results and some future directions.

While this paper was nearing completion, we received the interesting
work \HellermanNX, which has some overlap with the present note,
particularly in regards to the mostly harmless effects of
pseudotachyons, and which shows that a tachyon condensation process
in SCLDs for which the tachyon depends on a light-like coordinate
follows the RG flow even for small $Q$.

\newsec{Supercritical linear dilaton theories and the relation
of tachyon condensation to renormalization group flow}

The simplest non-trivial time-dependent background of string
theory is a linear dilaton background in which the dilaton is linear
in the time coordinate,
\eqn\lindil{\Phi = - Q X^0,\qquad g_s = e^{- Q X^0},}
with $Q > 0$ and with a flat string frame metric. We arbitrarily
choose the string coupling to decrease rather than to increase,
leading to a weakly coupled future asymptotic region in the
space-time.

As discussed above, systems such as this with weak coupling only in
the future or past asymptotia are generic, even in situations with a
lower scale of supersymmetry breaking (the background \lindil\ breaks
supersymmetry at the string scale).  For example, although type II
string theory on a Calabi-Yau manifold admits an exactly static
solution, and adding 3-form flux breaks the supersymmetry well below
the KK scale of the geometry at large radius, the flux changes the
behavior of the system in the far past to be strongly coupled and
infinitely far from this static solution.  With sufficiently generic
sources included in the system, the system may sit in a metastable
minimum of the effective potential immediately before tunneling into
the weakly coupled future asymptotic phase. This is a reasonably
natural way to ``cap off" the strong coupling regime, but still does
not fully account for the behavior arbitrarily far back into the
past. Another possibility is a transition from nothing, by tunneling
or tachyon dynamics.

The linear dilaton CFT has a central charge $c_{LD} = 1 - 3 Q^2$ (we
use conventions in which $\alpha'=1/2$). Together with any
(decoupled) unitary matter CFT with central charge
$c_{mat} = c_{crit} - c_{LD}$
(where for superstrings $c_{crit}=15$) it gives an exact
time-dependent background of classical string theory (of course, in
the past the coupling constant becomes strong so we need to embed
this into some complete quantum theory).  We will call these
theories supercritical linear dilaton theories (SCLDs)\foot{One can
think about these theories as theories with a large central charge
coupled to worldsheet gravity, where the worldsheet gravity gives
rise to a time-like Liouville field which we denote by $X^0$.
However, in this interpretation \refs{\PolyakovRD,\PolyakovRE}
one generally obtains also a worldsheet
Liouville potential for $X^0$, while we will assume that no such
potential is present, so our theory is a fine-tuned
deformation of the usual
coupling of a supercritical theory to gravity.}.

The linear dilaton significantly changes the behavior of
deformations of the worldsheet theory. The dimension of the operator
$e^{\kappa X^0}$ is given by
\eqn\dimkappa{{\rm dim}(e^{\kappa X^0}) = {1\over 4} \kappa
(\kappa + 2 Q),}
so a deformation of the worldsheet action by
\eqn\finaldeform{\int d^2z \sum_i \mu_i \cO_i e^{\kappa_i X^0},}
where the operator $\cO_i$ has dimension $\Delta_i$, is marginal when
\eqn\forkappa{\Delta_i + {1\over 4} \kappa_i (\kappa_i + 2 Q) = 2,}
or
\eqn\nforkappa{\kappa_i = -Q \pm \sqrt{Q^2 - 4 (\Delta_i - 2)}.}
The behavior of the deformation as a function of time depends on
$\Delta_i$. For relevant deformations of the ``matter CFT'', with
$\Delta_i < 2$ (or $m^2+\vec{k}^2 < 0$ where $m$ is the space-time
mass of the corresponding field and $\vec{k}$ is its spatial momentum
in any flat non-compact directions), one of the solutions for $\kappa$
is negative and one is positive, leading to a mode which grows with
time. This is similar to the usual case of tachyons in string
theory. For $2 <
\Delta_i < 2 + Q^2 / 4$ (or $0 < m^2+\vec{k}^2 < Q^2$),
both solutions for $\kappa$ in \nforkappa\
are real and negative. In this case the deformation decreases with
time, so it is not a tachyon; nevertheless such modes are unstable in
some senses that we will discuss below, so we will call them
``pseudotachyons''. Finally, for $\Delta_i > 2 + Q^2 / 4$ the
solutions for $\kappa$ are complex, and have a damped oscillatory
behavior at late times.

If we start from a matter theory with no relevant operators, the
background is stable, as no perturbations grow with time.  In this
background, modes are either oscillatory for all time, or
non-oscillatory for all time, and are not created or destroyed by the
time evolution \SCLDE, \gE.  The population of these modes depends on
the initial conditions coming from the strong coupling regime in the
far past.  Tunneling from a metastable de Sitter minimum, for example,
would imbue the system with a scale invariant spectrum of
fluctuations.

Various examples of such stable SCLD backgrounds can be constructed. For
instance, if we are in a superstring theory in which the matter
theory is a free theory of $9+16n$ superfields (in addition to the
fermionic partner of $X^0$), corresponding to superstring theory in
$d=10+16n$ dimensions, one can take a type II GSO projection and
obtain a modular-invariant partition function with no tachyons.
Obviously, the matter theory in such constructions does not have to
be a free theory (it could be, for example, a sigma model on a
Calabi-Yau manifold), so there is an infinite number of examples of
such stable theories (with various values of $n$). Presumably, there
are also many other possibilities for tachyon-free GSO projections
(such as heterotic theories).

If our matter theory has a relevant operator $\cO_i$, corresponding
to a tachyonic field in space-time, the theory is unstable toward
condensing this field, and it is interesting to ask what is the
end-point of this closed string tachyon condensation process, which
is described (at leading order in the deformation) by adding
\finaldeform\ (with the positive solution for $\kappa_i$) to the
worldsheet action. In general this question is very complicated.
However, it turns out that in the large $Q$ (or large $c_{mat}$)
limit the answer is simple -- the tachyon condensation process
precisely follows the renormalization group (RG) flow in the matter
CFT\foot{Note that both the tachyon condensation process and the
renormalization group flow are reversible in principle; however, in
general, reversing these processes requires significant fine-tuning
in order to go back exactly to the maximum of the tachyon potential
or to the UV fixed point of the RG flow.} corresponding to the
deformation by the operator $\cO_i$, with the RG scale proportional
to $e^{-2 X^0 / Q}$.  In the context of strings propagating on flat
space (or nearly flat space) this can be seen in two (equivalent)
ways \refs{\CooperVG\MukherjiJI\TseytlinYE-\SchmidhuberBV} (see also
\refs{\EllisEH,\EllisBA}). One way is by analyzing the beta function
equations of the deformed worldsheet action and noting that they are
equivalent to the matter RG equations to leading order in $1/Q$.
Alternatively, one can analyze the space-time equations of motion,
and note that the equation of motion of the tachyon $T$ (as of all
other NS-NS fields) contains a damping term coming from the linear
dilaton
%
\eqn\tachyoneom{{d^2 T \over (dX^0)^2} + 2 Q {dT\over dX^0} = -
{\del V \over {\del T}}.}
In the large $Q$ limit there is a solution to this equation in which
the first term on the left-hand side is negligible. This solution
has the tachyon
slowly rolling down its potential, and this is identified with the
RG flow of the matter theory. A more general argument (valid also
when the matter CFT is far from free) for this relation
between the tachyon condensation process and the RG flow is that
in the large $Q$ limit, the interactions induced by \finaldeform\
between the matter CFT and the $X^0$ CFT are of order $\kappa_i
\sim 2(2-\Delta_i)/Q$. Thus, at leading order in $1/Q$ there are
no interactions between the matter CFT and the $X^0$ CFT, and the
corrections to this picture scale as $1/Q$ \foot{Note that this
argument only applies to flows by relevant operators, not by
marginally relevant operators. Note also that another way to obtain
small values of $\kappa_i$ is by choosing $\Delta_i$ to be very close
to two. This also leads to a tachyon condensation process which begins
by following the RG flow, even when $Q$ is not large (see the recent
discussions in \refs{\FreedmanWX,\SuyamaWD}), but generally during the
flow $\Delta_i$ would deviate from two and the tachyon condensation
would deviate from the RG flow.}.

At first sight this picture does not make sense, since the
RG flow in the matter CFT leads to a smaller final central charge
$c_{mat,f} < c_{mat,i}$, while the full string theory must
remain critical also at late times after the tachyon has condensed (we
are assuming for simplicity that there is only a single tachyon, whose
condensation leads to a stable string theory). However, this change in
the central charge is compensated by a change in the linear dilaton
slope from $Q_i$ to $Q_f$, whose
magnitude in the large $Q$ limit is
\eqn\newq{3Q_{i}^2 - 3Q_{f}^2 = c_{mat,i}-c_{mat,f} \qquad \to \qquad
Q_{f} = Q_{i} - {{c_{mat,i} - c_{mat,f}} \over {6
Q_{i}}} + \cdots,}
which is of order $1/Q$ (compared to the difference between the
initial and final matter theories, which is of order one in the large
$Q$ limit)\foot{From the space-time point of view, the change in the
linear dilaton slope ensures that there is still a flat space solution
in the string frame even after the tachyon potential $V$ decreases
from a maximum to a minimum.}. Thus, it can be generated by the
interactions between the two sectors, as discussed in a set of
heterotic examples in \hellliu.

In the large $Q$ limit, the tachyon grows slowly:  $T\sim \mu
e^{X^0(2-\Delta_T)/Q}$.  This means that in this limit,
the string coupling decreases dramatically during a tachyon
condensation transition. When the tachyon increases from some initial
value $T_i$ to $T_f$, the string coupling changes by :
\eqn\Tslow{{g_{s,f}\over g_{s,i}}\sim \biggl({T_i\over
T_f}\biggr)^{Q^2/(2-\Delta_T)}.}
%
Note that this dramatic decrease in the string coupling during the
SCLD phase makes it difficult to obtain a transition to a realistic
FRW cosmology from a SCLD phase (even if the tachyon condensation
eventually leads to a critical background).

The relation described above between tachyon condensation and RG flow
is very general, but there is one important caveat; one must make sure
that the tachyon associated with the identity operator in the ``matter
CFT'' is not turned on, so this operator must be projected out by some
GSO-type projection both in the early-time and in the late-time
theory. When this ``ultimate tachyon'' is turned on, it seems to lead
to the end of space-time \McGreevyCI, instead of the flow we discussed
which ends with the IR matter CFT coupled to a time-like linear
dilaton. Except for this caveat, the relation is completely
general\foot{
At least for supersymmetric worldsheet theories, in which one
does not need to worry about the energy difference between
the vacuum energies of the initial and final matter theories.
}, and it seems that many RG flows (between
unitary theories with arbitrarily large central charges) can be
embedded in string theory in this way; the only requirement is that
one should be able to add to the matter theory we are interested in
some additional matter fields in a way which allows for appropriate
GSO projections that remove the ``ultimate tachyon''. For example, one
can discuss flows generated by mass terms for scalars (or by
sine-Liouville type interactions), which reduce the number of
dimensions of space (as do RG flows in linear sigma models). In some
cases (such as ${\cal N}=(4,4)$ linear sigma models) an RG flow can
lead to a sum of decoupled CFTs \refs{\AharonyTH,
\WittenYU}; in such cases the string theory at late times lives on
a sum of disconnected spaces (as in \AdamsRB) which could have
different dimensions.

The discussion above provides a large class of examples of closed
string tachyon condensation processes over which we have complete
control in classical string theory (assuming that we understand the
RG flow in the ``matter CFT'', that the corresponding operator is in
the NS-NS sector, and that the string coupling is already weak when
the tachyon starts condensing). So, one can use these theories to
ask various questions about closed string tachyon condensation, such
as how is the reduction in the high-energy density of states
implemented, which state does the vacuum of the initial theory
evolve into, how many particles are produced, etc. Of course, the
answers to these questions are not necessarily the same in the large
$Q$ limit as in the $Q=0$ case (where the tachyon condensation is
definitely not the same as RG flow), but still it is nice to have
controllable theories in which these questions can be answered. When
one tries to ask questions about quantum effects in the SCLD
theories, one immediately encounters a problem -- the one-loop
(torus) amplitude in these theories diverges.  In the next section
we will discuss the interpretation of this divergence, and how we
believe it can be resolved. In section 4 we will discuss the
particle production during the large $Q$ closed string tachyon
condensation process, and verify that it does not change the
classical picture of the time evolution described above.

\newsec{Pseudotachyons}

As mentioned above, the supercritical linear dilaton theory has a
spectrum of small perturbations including modes with
$m^2+\vec{k}^2<Q^2$ which do not oscillate. The deformation of the
worldsheet corresponding to such a mode $\eta$ decays exponentially,
but in space-time, the canonically normalized field $\tilde\eta$
behaves as
\eqn\canonorm{\tilde\eta\sim {\eta\over g_s}
\sim e^{QX^0}\eta\sim e^{\pm QX^0
\sqrt{1-(m^2+\vec{k}^2)/Q^2}}}
which can grow exponentially. From this point of view, these modes
might seem like tachyonic instabilities. However, even though these
modes grow exponentially, their back-reaction on the geometry, and
their self-interactions, are negligible because of the decreasing
string coupling $g_s\sim e^{-QX^0}$.  From the worldsheet point of
view, as discussed in \S2, these modes involve operators in the
matter CFT which are not relevant and do not decrease the effective
central charge.

In this section, we will characterize more generally this phenomenon
of {\it pseudotachyons} and its relation to the partition function
and the effective central charge. We begin in \S3.1 with a review of
the one-loop divergence of the SCLD partition function, and a general
discussion of tachyonic instabilities in time-dependent backgrounds.
In \S3.2 we discuss why in some cases (like SCLD backgrounds) these
instabilities are not deadly, and how to make sense of such backgrounds.


\subsec{1-loop IR divergences in SCLDs and other backgrounds}

We begin by computing the one-loop partition functions of SCLDs and
noting that the modes with $m^2+\vec{k}^2 < Q^2$ lead to an IR
divergence, as follows.
In order to discuss the one-loop partition function we first have to
address the divergence caused by the negative kinetic term for the
time-like field $X^0$.  Staying in Lorentzian space-time signature,
this may perhaps be accomplished (at least in the field theory
limit) by inserting a convergence factor $e^{-\epsilon \int
(X^0)^2}$; it would be interesting to analyze this in detail.
Another method which is available in some backgrounds (including the
SCLD background) is to perform a smooth Euclidean continuation which
renders the path integral manifestly convergent (up to IR effects
that are accessible in quantum field theory).
Since our worldsheet
Lagrangian includes a coupling of the form $Q X^0 R^{(2)}$, where
$R^{(2)}$ is the worldsheet curvature, it may be natural to Wick
rotate also $Q$ when we Wick rotate $X^0$; however, this does not
make sense since it would change the value of the central charge
(the total central charge would no longer vanish), so we will leave
$Q$ real.

In a critical limit of string theory in $d$ non-compact space-time
directions, the one-loop partition function per unit volume can be
expanded in the limit of a large imaginary part $\tau_2$ for the
modular parameter $\tau$ of the torus, and it takes the form
\eqn\oneloop{i \int^{\infty} {d \tau_2 \over {2 \tau_2}}
(2 \pi^2 \tau_2)^{-d/2} \sum_i e^{- \pi m_i^2 \tau_2 / 2}}
in terms of the masses $m_i$ of the string states. This diverges
exponentially when there are tachyons with $m_i^2 < 0$, but not
otherwise.

In the SCLD theory the exponent is shifted by $\pi Q^2 \tau_2 / 2$.
There are several ways to see this. From the point of view of
computing the partition function on the worldsheet, the presence
of the linear dilaton term has no effect on the torus, because we
can choose the worldsheet curvature to vanish everywhere. Thus, the
result is similar to that of the critical string theory, but with
the matter theory central charge bigger by $c_{mat} - c_{crit} + 1$.
Since the matter contribution to the integrand of \oneloop\
(from a specific state with a given worldsheet dimension) depends
on the central charge
as $e^{\pi c \tau_2 / 6}$, the integrand is multiplied compared
to the critical case by $e^{\pi (c_{mat} - c_{crit} + 1) \tau_2 / 6} =
e^{\pi Q^2 \tau_2 / 2}$.

Another way to see the same result is from the fact that the
worldsheet dimension of an operator with a specific frequency
is changed \dimkappa\ by $\kappa Q / 2$ from its original dimension.
Since the worldsheet path integral is a sum over $e^{-4\pi \tau_2
\Delta_i}$, the usual integral giving the contribution of a
particle with mass $m$,
\eqn\singleparticle{\int {d^d k \over (2\pi)^d} e^{-\pi \tau_2
(k^2 + m^2) / 2},}
which gives rise to the integrand of \oneloop, is replaced, using
\dimkappa, by
\eqn\nsingleparticle{
\int {d^{d-1} \vec{k} \over (2\pi)^{d-1}}
{d\omega \over 2\pi} e^{-\pi \tau_2 (\vec{k}^2 + m^2 + \omega
(\omega + 2 Q)) / 2} =
\int {d^{d-1} \vec{k} \over (2\pi)^{d-1}}
{d{\tilde \omega} \over 2\pi} e^{-\pi \tau_2 (\vec{k}^2 + m^2 + {\tilde
\omega}^2 - Q^2) / 2}}
in which all the masses are shifted by $m^2 \to m^2 - Q^2$, giving
the same result as before.

This means that any string state whose mass squared is smaller than
$Q^2$ (recall we are using units of $\alpha'=1/2$) behaves like a
tachyon in the sense of giving a divergence in the one-loop
partition function. Note that the states in the range $0 \leq m^2 <
Q^2$ are precisely the states which we called {\it pseudotachyons} in
the previous section; their zero momentum modes evolve as real
exponentials of the time variable, rather than having an oscillatory
behavior, and when canonically normalized they can grow
exponentially in time.  The IR divergence they produce is related by
modular invariance to the supercritical density of states,
as discussed in general in \kutsei, but as we
will discuss further below they do not produce a significant
instability.

Before moving to this interpretation, let us examine this effect in
greater generality. One-loop divergences of this type will occur
whenever we have fields whose effective mass squared (when they are
canonically normalized) becomes negative at late times, since this
leads to an exponential growth in these fields at late times.  Since
this is an infrared issue, we can describe the phenomenon in terms of
low energy effective quantum field theory and general relativity.
For simplicity we discuss the case of massless scalar fields (adding
mass and/or spin is straightforward). The one-loop amplitude in
general is of the form
\eqn\IRloop{\int d^d x\sqrt{-g}\Lambda(x)= \Tr \log({\cal H})=\int
{d\tau_2\over\tau_2}\Tr e^{-\pi \tau_2{\cal H} / 2}}
where ${\cal H}=\nabla^2$ is the worldline Hamiltonian (equivalently
the space-time Laplacian).
If in a complete basis of normalizable functions, the trace in
\IRloop\ diverges in the infrared, one has a tachyon or a
pseudotachyon.

The Feynman propagator contributing to perturbative amplitudes
satisfies
\eqn\genprop{\nabla^2 D = \delta (x-x')/\sqrt{-G}.}
Given a complete set of eigenfunctions of the Laplacian
$\{\Psi_n(x)\}$, satisfying
\eqn\Lappsi{\nabla^2 \Psi_n = \Delta_n\Psi_n, ~~~~~
\sum_n\Psi_n^*(x)\Psi_n(x')=\delta (x-x')/\sqrt{-G},}
we can construct the Feynman propagator as
\eqn\genFprop{D(x,x')=\sum_n{{\Psi_n^*(x)\Psi_n(x')}\over
{\Delta_n+i\epsilon}}.}
If $\Delta_n$ is negative for any $n$, then this produces an IR
divergence in \IRloop\
indicating the presence of a pseudotachyon or tachyon.

Let us examine this effect in a family of flat $n+1$ dimensional FRW
cosmologies
\eqn\FRWa{ds^2=-dt^2+a(t)^2d\vec x^2,}
with $H \equiv \dot{a}/a$ as usual.
Each Fourier mode with spatial momentum $\vec{k}$ of a massless
scalar field $\eta$ in this background
has an action
\eqn\etaac{S_{\eta}\sim \int dt
a(t)^{n}((\dot\eta)^2-{{\vec{k}^2} \over a(t)^{2}} \eta^2)=\int
dt(\dot{\tilde\eta}^2-m_{eff}^2(t)\tilde\eta^2),}
with $\tilde\eta$ the canonically normalized field and
\eqn\meff{m_{eff}^2(t)=-{n\over 2}(\dot H+{n\over 2}H^2)+{\vec{k}^2\over
a(t)^2}.}
In the inflationary limit of nearly constant $H$, there is a manifest
pseudotachyonic instability with the nearly constant negative mass
squared from the Hubble friction dominating over the gradient energy at
late times, leading to pseudotachyonic density perturbations.
Generically, this effective mass squared is time-dependent, but
whenever it becomes negative at late times we will get an
exponential growth of the canonically normalized field and
IR divergences.

To investigate this further, let us consider a single-component
homogeneous source, leading to a power law scale factor $a(t)\sim
t^\beta$,
\eqn\FRW{ds^2=-dt^2+t^{2\beta}d\vec x^2.}
In this case,
\eqn\meffbeta{m_{eff}^2={w\over {t^2(1+w)^2}} +  {\vec{k}^2 \over
t^{2\beta}},}
where $w\equiv (2-n\beta)/(n\beta)$ is the ratio of pressure to energy
density in the source.

For $\beta<1$, the gradient energy beats the Hubble friction
contribution at late times, and one might suspect from this that no
IR divergence results for these modes.  This is indeed correct, as can be
seen as follows.
The two-point function satisfies \genprop.  This is proportional to
\eqn\propFRW{ D(x,x')=\int d\vec k d\omega\omega
{{U_\omega^*(t)U_\omega(t') e^{i \vec x\cdot \vec
k}}\over{\omega^2+\vec k^2} },}
where
\eqn\Usoln{U_\omega(t)\sim t^{-\beta n/2} t^{1/2}
J_{(1-n\beta)/(2(1-\beta))}(\omega t^{1-\beta}/(1-\beta)). }
This expansion is in a complete set of modes since
%
\eqn\Jcomplete{ \int_0^\infty d\omega \omega J_\nu (\omega
t^{1-\beta}) J_\nu(\omega t'^{1-\beta}) =
\delta(t^{1-\beta}-t'^{1-\beta})/t^{1-\beta}.}
Thus, for $\beta<1$, there is a complete set of modes for which the
denominator in the expression \genFprop\ is never negative, so we do
not see the kind of divergence present in (pseudo-)tachyonic
theories.  The case $\beta=1$ arises in the SCLD phase analyzed
above, giving pseudotachyonic modes as discussed there. Note that
within this family of solutions \FRW, we find no pseudotachyons for
a decelerating scale factor, but for $\ddot a\ge 0$ the Hubble
friction competes with the gradient energy and pseudotachyons
appear.

Having discussed the issue in some generality, let us make one final
remark. In cases where it occurs in a weakly coupled string theory,
an IR divergence in the partition
function is related by a modular transformation to a supercritical
effective central charge, as explained for example in \kutsei
\foot{As we will discuss in the next subsection, this
IR divergence does not necessarily imply a catastrophic instability on
equal footing with that in the bosonic string theory.}.
In the SCLD case of interest here,
$c_{eff}=3 Q^2+c_{crit}$ is part of the construction.  However,
pseudotachyons also appear in the infrared in many other late-time
cosmological solutions. This leads to a large class of interesting
backgrounds where the supercritical density of states arises in
novel ways, as in compact negatively curved spaces where it arises
from winding strings supported by the fundamental group
\refs{\mutation, \growthnotes}.

\subsec{Pseudotachyons and their resolution}

At first sight the divergence we found above in the one-loop
partition function implies that the theory is ill-defined, just like
any other theory containing tachyons, and that we are (in some
sense) expanding around an unstable vacuum of the theory. However,
we find that this is not correct as a general statement:  many
backgrounds, such as the SCLD theory, are not significantly affected
by the condensation of the pseudotachyonic modes.
The real problem with tachyons is not the fact that they grow
exponentially with time, but that their back-reaction on the
original configuration grows exponentially with time, so that
including this back-reaction moves us far from our original starting
point (which is why it does not make sense to expand around this
original starting point). This is not the case for the
pseudotachyons in the SCLD background. Since all the interactions of
the pseudotachyons with other fields come with factors of $g_s =
e^{-Q X^0}$ (when they are canonically normalized), their
back-reaction on other fields (which is proportional to $g_s$ times
the pseudotachyon field) actually decays exponentially with time,
unlike that of standard tachyons (with $m^2 < 0$). Thus, we claim
that even though the one-loop partition function of SCLDs diverges
because there is an instability toward producing
exponentially-growing (canonically normalized) pseudotachyon fields,
this instability does not take us away from our original SCLD
configuration, so this configuration is actually stable. A similar
claim can be made about other cosmological backgrounds of the type
described above, as long as all the coupling constants decay fast
enough. As is the case for perturbations of light fields in the
inflationary universe, we expect that the expectation values
$\vev{{\tilde \eta}^2}$ of the squares of the canonically normalized
pseudotachyon
fields will grow with time, as in a random walk process, and these
fields will be in a highly squeezed state. In the standard
inflationary case the perturbations eventually decohere, but it
seems that in the SCLD background this will not happen because of
the exponentially decaying interactions; however, this difference
between these two examples of pseudotachyons will not be important
for our discussion here.

Note that zero momentum massless string modes, or moduli, do affect
the background at late times so it is important to keep track of
them (if they exist) in SCLD backgrounds (as in standard string
backgrounds). These modes would shift by a finite amount
during the closed string tachyon condensation process discussed
in the previous section. Even when a
potential for these modes is generated at string loops, the friction
and the exponential decay in the potential generally prevent them
from going to the minimum of this potential, so these moduli really
are parameters of the SCLD background even though it is not
supersymmetric.  To see this, consider the equation of motion for
such a rolling scalar field $\phi$ in the presence of a 1-loop
tadpole $V_1$
(taken to be a constant for simplicity), which is of the form
\eqn\modeom{{{d^2\phi} \over{(d X^0)^2}} + 2 Q{{d\phi}\over dX^0} = -
e^{-2QX^0} V_1.}
The solution takes the form
\eqn\modsol{
\phi=\phi_0+\phi_1e^{-2Q X^0}+e^{-2QX^0}{V_1X^0\over {2Q}}.
}
This shows that the field at late
times reaches a constant, which can take different values $\phi_0$.

Although it is interesting to monitor the evolution of the moduli
during the pseudotachyon condensation process, it is also sometimes
useful to consider a weaker notion of stability, namely to ask
whether the pseudotachyon mode changes the effective central charge
of the background.  This measure of stability does not depend on
where the system ends up on its effective moduli space. It simply
measures whether the process lifts enough worldsheet degrees of
freedom to modify the leading Hagedorn density of states. Tachyons
built from relevant operators in the worldsheet matter CFT do change
this quantity; $c_{eff}$ decreases in their condensation process. In
space-time this corresponds to a process in which masses are
increased by the tachyon condensation
\refs{\StromingerPC,\StromTak,\McGreevyCI}. As we have seen here,
modes built from marginal or irrelevant operators in the matter CFT
on the worldsheet do not change $c_{eff}$, at least in the SCLD case.

Therefore, we claim that SCLDs are consistent stable string
backgrounds, even when the loop corrections are taken into account.
The divergence in the one-loop partition function arises because
there is an instability toward creating the exponentially growing
pseudotachyon fields, and in our naive partition function above we
got the divergence by summing over all possible values of these
fields (since larger values lead to a lower action). However, in the
actual physical situation that we are interested in, these fields
will be in some initial state which depends on the initial
conditions (in any case, in order to really be able to sensibly
discuss loop corrections, we need to assume that the strong coupling
region at early times has been capped and replaced by some other
background). This initial state could be a classical state which
would sit at some distance away from the origin of field space and
then exponentially grow from there, or a quantum state which could
be a superposition of such states. However, for a given initial
state we should not sum over all possible values of the
pseudotachyon fields, so we would not encounter the IR divergence.
Since the back-reaction of the pseudotachyons is small, the details
of the state they are in do not affect the fact that the future will
be well-described by the SCLD background \lindil. They do, however,
affect various measurements performed in this background, like the
one-point or two-point functions of the pseudotachyon fields. Thus,
we cannot predict the results of such measurements without knowing
the initial state. Nevertheless, this does not affect the stability
of the SCLD background.

One specific initial state which is particularly simple is the one
defined by analytically continuing $Q \to i Q$ (and then continuing
back to obtain the final answer). As discussed above, it is not
clear how to implement this directly in string theory, but (since
the pseudotachyon divergence comes from the IR) we can implement it
at the level of the low-energy effective action. For modes which are
not pseudotachyonic, with $\vec{k}^2 +m^2 > Q^2$, the one-loop
partition function may be written as a sum of the zero-point
energies $\sqrt{\vec{k}^2 + m^2 - Q^2}$. For the pseudotachyonic
modes, we can use the same expression, defined by the analytic
continuation to be purely imaginary and equal to $i \sqrt{Q^2 - m^2
- \vec{k}^2}$. Essentially, we are taking the contribution of every
mode corresponding to a negative harmonic oscillator to be the
(imaginary) frequency of this oscillator; this is related to the
instability of this oscillator towards generating an exponential
growth of the field at this rate. Using this prescription, we find
that the contribution of a field of mass $m$ to the one-loop
partition function takes the form $|m^2 - Q^2|^{d/2} \log(m^2 -
Q^2)$, which agrees with the standard Coleman-Weinberg answer when
$m^2 > Q^2$ but has an imaginary part (described above) otherwise.
Note that this formula is for each string mode separately, so it
should be multiplied by a Hagedorn density of states. As we
mentioned above, it is not clear how to perform this analytic
continuation directly in the worldsheet action; we believe that a
similar continuation can make all higher-loop corrections to SCLD
computations well-defined, but we do not know how to do this
explicitly, and it would be interesting to verify this. In any case,
we stress that this continuation corresponds to a specific choice of
the initial state, and that most initial states will lead to
different results. Presumably, the description of generic initial
states on the worldsheet would be non-local \AharonyCX\ since they
would look like squeezed states.

\newsec{Particle production}

In section 2 we described how in the large $Q$ limit, tachyon
condensation in SCLDs followed the RG flow on the worldsheet. The
arguments for this were based on a classical analysis, which ignored
the particles which are produced quantum-mechanically in
time-dependent backgrounds. It is interesting to ask how many
particles are produced during the tachyon condensation process, and
whether these particles change the claim that the end-point of the
process is another SCLD theory with smaller $Q$ or not.

As described in section 2, the SCLD background and the tachyon
condensation have a simple description on the worldsheet. However,
it seems difficult to analyze the particle production directly on
the worldsheet, which requires computing 2-point (and higher)
correlation functions in the time-dependent CFT (including the
tachyon operator \finaldeform) of section 2.
Instead, we will estimate the particle production by computations
done in the space-time effective field theory. A priori one would
not expect to be able to trust such an effective field theory in a
background in which the dilaton changes at a very fast pace
(compared to the string scale).  However, the linear dilaton is an
exact classical solution.  It introduces friction on other modes
such as the tachyon, and the light modes change
slowly compared to the string
scale.  We consider the physics at sufficiently late times that the
coupling is very weak, and self-consistently work to linearized
order in all other fields.


As we discussed above, only particle modes with $m^2 + {\vec k}^2 >
Q^2$ oscillate in the SCLD background, so they can be given the
standard particle interpretation and we can ask about their
production\foot{In this section we ignore the fact that $Q$ changes
with time, since this is small in the large $Q$ limit as discussed
in \S2.}. Let us begin with a simple toy model -- the particle
production of such a mode $\Phi$ (with mass $m$ and momentum ${\vec
k}$) when it is coupled to the tachyon $T$ by a ${1\over
{2g_s^2}}\lambda T^2 \Phi^2$ potential in the string-frame space-time
effective action. In the presence of the condensing tachyon, going
as $T = \mu e^{\kappa X^0}$, we obtain an equation of motion
\eqn\phieom{{d^2 \Phi \over {(d X^0)^2}} + 2 Q {d\Phi \over {dX^0}} +
(m^2 + {\vec k}^2 + \lambda \mu^2 e^{2 \kappa X^0}) \Phi = 0.}
It is easy to find the general solution to this equation, which is a
generalization of the solution for $Q=0$ discussed in \refs{
\BirrellIX,\StromingerPC,\McGreevyCI}. It may be written as a sum
of two Bessel functions in the form
\eqn\phisol{\Phi = \alpha e^{- Q X^0} J_{{\sqrt{Q^2-m^2-\vec{k}^2}
\over \kappa}}\left({\sqrt{\lambda} \mu \over \kappa} e^{\kappa
X^0}\right) + \beta e^{- Q X^0} J_{-{\sqrt{Q^2-m^2-\vec{k}^2} \over
\kappa}}\left({\sqrt{\lambda} \mu \over \kappa} e^{\kappa
X^0}\right).}
When $m^2 + \vec{k}^2 > Q^2$, the phase of these solutions oscillates
both at very early and at very late times, and a standard computation
of the Bogolubov coefficients gives the average number of created
particles, which is $\exp(-\pi \sqrt{m^2 + \vec{k}^2 - Q^2} /
\kappa)$.  For large $Q$, when $\kappa$ is small (of order $1/Q$), and
when the momentum or the mass of $\Phi$ are large enough, this
particle production is very small. Note that in the large $Q$ SCLD
vacua (at weak coupling), most of these produced particles would not
decay (despite the exponentially growing phase space factor which is
available to them due to their exponentially growing mass) because of
the rapid exponential decrease in the coupling constant.

If we consider the case of several particles with more complicated
couplings to the tachyon which mix them together, we can no longer
exactly solve the equation of motion as above. However, in the limit
in which all particles have $m^2 + \vec{k}^2 - Q^2 \gg \kappa^2$, the
time evolution is very slow (assuming polynomial couplings to the
tachyon), and we can use the adiabatic argument (as presented in
\BirrellIX) to prove that the particle production rate is smaller than
any power of $\kappa^2 / (m^2 + \vec{k}^2 - Q^2)$, suggesting that it
is still exponentially small as in the explicit computation above.
This argument can be used both for the production of particles whose
mass eventually goes to infinity, as in \phieom\ (this
is the case for most particles
when the central charge of the matter theory decreases), and for the
production of particles whose mass remains finite.

For large $Q$, as discussed in the previous section, many of the
string modes (with low momentum) are actually pseudotachyons which
do not propagate (their wavefunctions do not oscillate). During the
tachyon condensation, such pseudotachyons (or linear combinations of
pseudotachyons and regular particles) can develop exponentially
growing masses as above, and start propagating; this is somewhat
similar to what happens to modes which start outside the horizon in
an expanding decelerating FRW universe\foot{ Modes could also become
propagating even without interacting with the tachyon, just because
of the decrease in the value of $Q$. However, this effect can
be neglected in the large $Q$ limit.}. In this case there is no
exponential suppression of the particle production of the type
discussed above.

The precise details of the particle production in this case depend
strongly on the initial state of the pseudotachyons; if we compute
it by analytical continuation in $Q$ we find a particle production
of order one. More generally,
we can estimate the energy density contained in these modes as
follows. During the SCLD phase, while their effective mass is still
negative, these fields roll down their
effective potential $-m_{eff}^2\tilde\eta^2$ \meff\ by a
distance in field space $\tilde\eta_0$ depending on their initial
value at the beginning of the SCLD phase, and on the time spent in
the SCLD phase before the tachyon turns over their mass squared.
Once the tachyon dominates, these modes roll back toward the origin
and oscillate about the minimum, with an energy density of order
$\lambda \mu^2 e^{2\kappa X^0}\tilde\eta_0^2$ for each mode. This energy
density grows exponentially with time, and on top of it there is a
Hagedorn density of such modes, that is a density of states of mass
$m$ of order $e^{\pi m Q \sqrt{2}}$, for $m_s\ll m\le Q$.

Despite this large density of states, the energy density is finite
and we can begin with a sufficiently small string coupling such that the
process is under control throughout. The back-reaction of these
modes is down by a factor $g_s^2\sim g_{s,0}^2 e^{-2QX^0}$, which
guarantees that their effect is negligible (except when we
directly ask questions about these particles).


To summarize, in the large $Q$ limit of the tachyon condensation
process we expect to have a large production of particles which
start out as pseudotachyons with $m^2 + \vec{k}^2 < Q^2$ but become
propagating degrees of freedom, both for the particles which become
infinitely massive and for the particles which remain of finite
mass. Particles which begin with a larger mass or momentum are
produced only in small numbers. This large particle production
creates an exponentially large density of particles, and the mass of
most of these particles grows as a power of $e^{\kappa X^0}$
\foot{Note that this growth is in units of the string scale; the
mass decreases very fast in Planck units.}. However, since the
string coupling decays as $e^{- Q X^0}$, the back-reaction of this
large energy density decays very fast with time (note that the total
energy density diverges in the large $Q$ limit, but it is finite for
a given large value of $Q$). Thus, we believe that the large quantum
creation of particles during the tachyon condensation process does
not change the end-point of this process which we described in
section 2.

\newsec{Summary and future directions}

In this note we discussed SCLD backgrounds of string theories and
related backgrounds. We showed that despite the naive instability
of these backgrounds, related to the divergence of the one-loop
partition function, these backgrounds are actually well-defined
(assuming that they start from an appropriate initial state). We
reviewed the fact that in the large $Q$ limit of SCLD backgrounds,
the closed string tachyon condensation process on the worldsheet
follows the RG flow of the worldsheet CFT, and we argued that this
is not changed by quantum effects despite the fact that these
effects lead to a very large particle production in space-time.

It would be interesting to analyze further the possible initial
``capping'' states for SCLD backgrounds. One could imagine
smooth cappings involving either some strong coupling completion of
string theory (if this exists for supercritical strings), or a
smooth evolution from ``nothing'' as in \McGreevyCI. Alternatively,
the capping could involve a tunneling from some meta-stable background
of string theory (or perhaps from nothing). Given a specific
capping, one can compute the initial state of the SCLD phase, and
use this to predict the details of the final state of this phase.

Our stability analysis in this paper was purely perturbative, and
it would be interesting to analyze whether SCLD backgrounds are
stable also with respect to tunneling into other backgrounds or not.
Naively it seems that they should be, since the rapidly decreasing
coupling constant should suppress all tunneling amplitudes. If these
backgrounds are completely stable, then they provide an infinite
class of possible non-supersymmetric
end-points for evolution in the landscape (which
in some cases are labeled by moduli). It would be interesting to try
to estimate the probability of an eternal inflation process ending up
at such a background, as compared to the probability of a more
conventional string background.

The SCLD backgrounds provide controlled examples of time-dependent
backgrounds, and it is interesting to ask if they could have any
cosmological applications. One interesting point that was noted
already in \AntoniadisUU\ is that the SCLD backgrounds can solve
the horizon and flatness problems, so they might provide an
alternative to inflation. However, as mentioned above, the rapid
decrease in the string coupling in the SCLD backgrounds seems to
prevent a smooth transition from such backgrounds to our current
universe.

In this paper we focused on discussing the tachyon condensation
processes in the limit of large $Q$, in which one can rigorously
argue that the tachyon condensation process follows the RG flow
on the worldsheet. One may conjecture that also for smaller values
of $Q$, when the tachyon condensation process deviates from the
RG flow, it could still end up at the same end-point as the
RG flow, as long as $Q_f$ in \newq\ is positive (or maybe even
vanishing). However, for small $Q$ our arguments suggest that the
quantum effects are very important, so it does not seem possible
to check this conjecture just by using classical string theory.
It would be interesting to investigate what can be said about
tachyon condensation processes when $Q$ is not very large, and
in particular when the final $Q$ vanishes. This is necessary
in order to check the conjectures made in
\refs{\HellermanZM,\hellliu}.

\bigskip

\centerline{\bf Acknowledgements}

We would like to thank S. Hellerman, D. Kutasov, A. Maloney, J.
McGreevy, J. Polchinski, D. Starr, and I. Swanson for useful
discussions. O.A. would like to thank Stanford University, SLAC,
Cambridge University and the Aspen Center for Physics for
hospitality during the course of this work. The work of O.A. was
supported in part by the Israel-U.S. Binational Science Foundation,
by the Israel Science Foundation, by the European network
MRTN-CT-2004-512194, by a grant from the G.I.F., the German-Israeli
Foundation for Scientific Research and Development, by Minerva, by a
grant of DIP (H.52), and at the Institute for Advanced Study by the
DOE under contract DE-FG02-90ER40542. E.S. is supported in part by
the DOE under contract DE-AC03-76SF00515, by the NSF under contract
9870115, and by an FQXI grant, and thanks the Weizmann Institute and
KITP for support during the course of this work.

\listrefs

\end